\documentclass{article}

\usepackage{cite}
\usepackage{graphicx}
\usepackage{color,soul}
\PassOptionsToPackage{hyphens}{url}
\usepackage{hyperref}
\usepackage{booktabs} \usepackage{listings}
\usepackage{multirow}
\usepackage{mdframed}
\usepackage{bm}
\usepackage[accepted]{sysml2019}

\begin{document}
\graphicspath{{./}{figures/}}

\interfootnotelinepenalty=10000
\twocolumn[
\sysmltitle{Hoard: A Distributed Data Caching System to Accelerate Deep Learning Training on the Cloud}

\begin{sysmlauthorlist}
	\sysmlauthor{Christian Pinto}{irl}
	\sysmlauthor{Yiannis Gkoufas}{irl}
	\sysmlauthor{Andrea Reale}{irl}
	\sysmlauthor{Seetharami Seelam}{ytown}
	\sysmlauthor{Steven Eliuk}{svalley}
\end{sysmlauthorlist}

\sysmlaffiliation{irl}{IBM Research, Dublin, Ireland}
\sysmlaffiliation{ytown}{IBM Research, Yorktown Heights, NY, US}
\sysmlaffiliation{svalley}{IBM Silicon Valley Lab, San Jose, CA, US}

\sysmlcorrespondingauthor{Christian Pinto}{christian.pinto@ibm.com}
\sysmlcorrespondingauthor{Andrea Reale}{realean2@ie.ibm.com}

\sysmlkeywords{Machine Learning}

\vskip 0.3in

\begin{abstract}

Deep Learning system architects strive to design a balanced system where the computational
accelerator --  FPGA, GPU, etc, is not starved for data. Feeding training data fast 
enough to effectively keep the accelerator utilization high is difficult when utilizing dedicated 
hardware like GPUs. As accelerators are getting faster, the storage media \& data buses 
feeding the data have not kept pace and the ever increasing size of training data further 
compounds the problem.

We describe the design and implementation of a distributed caching system called Hoard
that stripes the data across fast local disks of multiple GPU nodes
using a distributed file system that efficiently feeds the data to ensure minimal degradation in 
GPU utilization due to I/O starvation. Hoard can cache the data from a central storage system before the
start of the job or during the initial execution of the job and feeds the cached data for
subsequent epochs of the same job and for different invocations of the jobs that share the same data requirements, 
e.g. hyper-parameter tuning. Hoard exposes a POSIX file 
system interface so the existing deep learning frameworks can take advantage of the cache without any modifications.
We show that Hoard, using two NVMe disks per node and a distributed file system for caching, achieves a
2.1x speed-up over a 10Gb/s NFS central storage system on a 16 GPU (4 nodes, 4 GPUs per node) cluster
for a challenging AlexNet ImageNet image classification benchmark with 150GB of input dataset.
As a result of the caching, Hoard eliminates the I/O bottlenecks introduced by the shared storage 
and increases the utilization of the system by 2x compared to using the shared storage without the cache. 
\end{abstract}
]

\printAffiliationsAndNotice{}

\section{Introduction}

Recent advancements in artificial intelligence are fueled by deep learning
techniques. Deep Learning (DL) is a class of machine learning (ML) techniques
that achieved notable success in speech recognition, visual recognition, and
language understanding. This success is due to three main advancements:
availability of massive amounts of data, commodity accelerator hardware
that can process this data faster such as FPGA, and GPUs, and the advancements in the neural
network models and programming frameworks.

DL system architects strive to design a
balanced system where the computational accelerators, e.g., GPUs, can achieve high utilization.
While the GPUs provide the computational capability, data is the fuel that
propels the DL training operations and keeps these GPUs busy. 
With growing
training dataset sizes, especially in video and image processing,
and increasing GPU speeds with every new
generation of GPUs \cite{nvidia-p100,nvidia-v100}, feeding the training data fast enough to keep
the GPUs busy is an increasingly difficult problem. There are two popular design patterns to overcome this
problem but both of them have shortcomings.

In the first design pattern, DL systems are built with high performance
storage, e.g. Solid State Drives (SSDs), or Non-Volatile Memory express (NVMe)
drives on the GPU nodes \cite{bhattacharjee2017ibm, mapr, aws, paperspace, nvidia-dgx, ibm-cloud}.
On these systems, users copy the training data from central storage systems, e.g.
 NFS, GPFS, Cloud Object Store into these high performance drives to feed
the GPUs more efficiently. This is the most common approach we find today on-prem and in the cloud.
In this model, additional compute nodes can be added incrementally, for example by renting more servers on the cloud,
and newer hardware such a recent versions of GPU or CPU systems can be added to the DL system. 
It works well for many cases but its scalability suffers from the dependency on the central storage
systems and the drives on the nodes pose operational problems.

There are several problems with this approach: 
the data is copied from the central storage into these local disks and deleted as soon as the job is terminated to make room for 
new jobs. 
As a result, the data is copied at the start of every job and in large hyper-parameter tuning 
experiments, where tens to hundreds of parallel jobs are started on different nodes, the data copy is extremely 
taxing on the shared storage servers. 
In addition, these high performance storage drives on the nodes tend to be limited in capacity (typically ~1TB) 
so jobs with large datasets fail to run because the data does not fit on the disks.
Since the GPU systems, in general,
are expensive and consist of multiple GPUs per node, they are often multi-tenant 
where storage and compute time is shared.
In space sharing, different GPUs of the node are allocated 
to different user jobs and the data for all the jobs running on the node must fit in the limited capacity
of the underlying storage on the node. When one job takes a subset of the GPUs on the node but takes up
most of the disk space, other jobs taking free GPUs on the node, fail to start because their data cannot be copied to the disk.
In any case, every time a job starts, its data must be copied which could take many hours depending 
on the size of the input dataset, the network bandwidth, contention on the network and the disk bandwidth.
This gets worse when a user
needs to use more than one system to train a model with multiple GPUs using a data
parallel approach, most common approach in DL training. In a data parallel approach, DL training
on each node typically accesses all of the dataset, albeit in random order, but multiple times
during the training process. This requires that each node has the entire dataset to train the model.
For these reasons, the high performance drives on different nodes hold copies of the entire dataset, 
which is not an efficient use of their limited capacity.

In the second design pattern, DL systems are built with high performance computing (HPC) like
dedicated storage systems connected to GPU clusters with high-end networking
gear such as Infiniband or 100Gbps Ethernet \cite{netapp, mapr, zou, ccc}. 
In this design, a small set of GPU nodes (typically $<$10) are connected to a dedicated storage server (see
\cite{netapp} as an example).  While this design 
addresses the I/O bottleneck problem and keeps the GPUs busy, and it supports
larger datasets, it is expensive to scale such a solution to thousands of users and
hundreds of servers with thousands of GPUs. This solution also does not give the flexibility to 
expand the cluster capability incrementally based on the workload and user demand, and 
seamlessly transition to newer hardware such as new versions of the GPUs and CPUs.

In this paper, we propose Hoard, a distributed caching system to address the limitations of the 
first design pattern for DL training on cloud environments. Hoard takes advantage of the 
unique access patterns of large-scale deep learning training and it is designed to address the needs of
a  multi-user, multi-tenant large scale training environment.
The main contributions include:
\begin{itemize}

   \item \textit{Design of a distributed cache management system that uses the fast local disks of compute nodes
and stripes the job's data across a subset of nodes in a configurable fashion}. Data striping
and holding data across multiple disks on multiple nodes allows the system to feed the GPUs
at the rate they can consume the data. We use a distributed file system for the data caching 
after careful evaluation of a number of them against the requirements for DL workloads. 
Although we use a particular file system, the implementation is flexible enough to integrate a different file system backend
for the distributed caching logic. By aggregating the space across multiple nodes, users can run jobs
with much larger datasets than those that fit in the disks of a single node. 

   \item \textit{Development of a simplified usage model so users can create cache objects that refer to their 
datasets in remote storage such as NFS and Object store and control their life cycle such a prefetch into 
the cache or delete from the cache independent of the job life cycle}. This takes DL developer training work-flow
into account. This model allows for users to cache datasets and run hyper-parameter experiments using that data without the overhead of 
copying the data for each job and for each hyper-parameter experiment. 

   \item \textit{Demonstrating that Hoard can achieve the same performance as the local disk, that it adds
minimal overhead, it achieves at least 2x speed-up over jobs that access data from remote storage and because
of Hoard, the cluster can support 2x more jobs without taxing the shared storage system}. Since the number of 
local disks scale with the number of nodes, Hoard can cache more data as more nodes are added to the cluster
and as a result it can support larger datasets with the increasing scale of the system. As systems get larger, aggregate throughput
of the system increases so the performance efficiency of cluster to run multiple jobs increases over using a shared
storage system.

\end{itemize}

The problems we discussed with the DL system architectures and our results suggest that 
deep learning workload and developer work-flow  aware distributed caches that leverage compute node local storage, memory and 
more modern memories like storage class memories (SCM) are necessary to ensure that the next generation
faster accelerators can be feed with the data necessary for them to make forward progress. Such cache systems
allow for better scalability of the DL systems on commodity hardware, more so in public cloud environments,  without
the need for HPC like storage and networking solutions.

The rest of the paper is organized as follows: Section 2 further describes the challenges with the first 
DL system design pattern and presents the requirements that drove the design of Hoard. Section 3 describes the 
user experience of Hoard and the underlying design of the caching system. Section 4 presents a comprehensive performance
evaluation of Hoard across multiple dimensions. We present related work in Section 5 and describe our conclusions and
future work in Section 6.
 
\newmdenv[leftline=false,rightline=false]{topbot}

\section{Requirements for large-scale deep learning data acceleration}
\label{sec:requirements}
Caching ``far'' data to closer and faster storage is, per se, an idea
probably as old as computer science \cite{nelson1987caching}.
For what concerns file system data,
which is the type of data we are concerned with in this work, most
modern operating systems transparently use free main memory to cache
``hot'' file blocks (e.g., the buffer-cache mechanism in Linux
\cite{buffer-cache}).  We argue, however, that existing mechanisms do not match
the requirements imposed by the unique access patterns of large-scale
deep learning training.

First, the size of training datasets can easily exceed that of any
individual server's main memory; although it is not uncommon to see high
end servers with few TBs of RAM but such servers are not common in cloud environments. 
Also, the amount of training data is deemed to
increase at a much faster pace (some open datasets are already over the TB
barrier \cite{openimages}, with private datasets possibly
growing even faster); increasing the amount of server memory at the same
pace is clearly not economically viable, also considering that memory
dedicated to caching is usually only a portion of the total server memory.
Caching on local secondary storage (e.g., NVMes, SSDs or spinning disks -
each with its own speed/cost trade-offs) is a more viable solution
\cite{bent2002flexibility,makatos2010using,li2014nitro,
Saxena:2012:FLC:2168836.2168863,byan2012mercury}.
However, in the increasingly common
case where entire datasets must be accessed by a training job in each node,
for example, in the case of large distributed training or parallel
model hyper-parameterization, replicating the whole dataset in each
of the servers' secondary storage cache is often infeasible because of size of data, costs, and can 
lead to quick exhaustion of cluster storage capacity. This drove our first requirement. 

\begin{topbot}
	\textbf{Requirement 1}: \emph{The cache should be implemented as secondary
	storage-based distributed cache, and it must effectively leverage the aggregate
	storage capacity across a subset of the nodes.}
\end{topbot}
This means that the dataset can be as big as the aggregate secondary storage of the entire DL system so
the training job is no longer limited by the size of the secondary storage on a single node. 

Second, existing cache systems normally consider the file or file-block
as the unit of cache management granularity. For example, the Linux
buffer-cache adopts a least-recently-used (LRU) cache replacement policy
that keeps in cache the most recently accessed file blocks while evicting
the oldest ones. This type of granularity does not fit the access patterns
of deep learning training jobs. In fact, every epoch of a training job
accesses the full dataset~\footnote{at least statistically, in case of
random batch sampling}; this means that, in case of cache contention,
evicting a fraction (only some files or file-blocks) of a dataset is
as good as not having any part of the dataset in cache. This is because
the next training epochs will have to access again the fraction of the dataset
that was previously evicted, leading to further evictions and, fundamentally,
to cache trashing effects. This leads to our second requirement:

\begin{topbot}
	\textbf{Requirement 2}: \emph{Cache management policies should operate at
	the granularity of the dataset. The life cycle of the dataset in the cache is 
decoupled from the life cycle of the job. A dataset may be in cache well after a job has completed the execution.}
\end{topbot}
This ensures that the dataset is in the system so repeated executions of the job with think times (typically developers
train a model for some time, observe the convergence, kill the job, restart the training with a different set of parameters and repeat)
and hyper-parameter training where lots of jobs execute in parallel benefit from the cached data.

In a distributed cache, every training job will access data
from possibly different nodes in the data-center. Accessing non-local
data could have two related and non-negligible consequences: (i) non-local data
access could be slower than local-data access and, most importantly, (ii)
non-local data access will use part of the capacity of the data-center
network to move cached data across servers. While state-of-the art
data-center networks (e.g., 40 GbE and 100 GbE networks are increasingly
common)  arguably provide more bandwidth than modern GPUs can consume
while training typical deep learning models (see Section 4), 
we expect that new deep learning accelerators (either new
GPUs or special purpose chips) will soon push this limit. Also, since
the data network is not dedicated to the distributed cache,
and it is used for application-level traffic (communication, synchronization) as well, 
this cache usage may result in impacting the communication dominated applications.
This results in our third requirement:

\begin{topbot}
	\textbf{Requirement 3}: \emph{Caching and job scheduling should be done in
	synergy. The job scheduler should co-locate cached data and training jobs
	taking into account the data-center network-topology to maximize access speed
	and minimize interference to application performance.}
\end{topbot}

Lastly, we want
Hoard to be fully transparent and agnostic to the
specific deep learning framework used to program the training jobs (e.g.,
TensorFlow \cite{tensorflow2015-whitepaper} or Caffe \cite{jia2014caffe}) so it could be
used with any DL framework.

\begin{topbot}
	\textbf{Requirement 4}: \emph{Cached data should be exposed transparently
	using a POSIX-compatible file system interface.}
\end{topbot}

The next section describes the architecture of Hoard and its implementation to satisfy the above four requirements.
 
\lstset{basicstyle=\footnotesize\ttfamily,breaklines=true, numbers=left,
xleftmargin=2em,framexleftmargin=1.5em, captionpos=b, frame=tb}

\section{Hoard Deep Learning Data Accelerator Architecture}
Hoard assumes that there is a resource manager 
that can deploy and execute deep learning jobs and a distributed
caching layer to create and delete the distributed data objects
that correspond to user datasets.
The architecture of Hoard with its components are shown in
\figurename{}~\ref{fig:impl-arch}. 
 Hoard is designed as a collection
of micro-services with clearly defined interfaces and well defined
functional boundaries that interact with the scheduling layer and the caching layer. 
For our implementation, Hoard assumes that deep learning jobs are deployed as
containers and that the available cluster resources are scheduled by the
Kubernetes \cite{kube-website} container orchestration system. It could be 
easily implemented on top of cluster resource schedulers like Mesos, Yarn,
and IBM Spectrum Conductor. We will discuss the technology selected for 
the distributed caching layer in Section 3.3.

In the following subsections we provide an overview
of how Hoard is designed and implemented by first introducing
the desired end-user experience (Section~\ref{sec:ux}) before
documenting how this behavior is realized on our current
implementation (Section~\ref{sec:impl}). 

\subsection{User experience}
\label{sec:ux}
As shown in \figurename{~\ref{fig:impl-arch}}, \textit{Hoard API Server} exposes two sets of APIs
that provide two main functions. One set of APIs allows the user to create new
datasets, query the cached datasets, and delete the cached datasets. The second
set of APIs allows the user to tie the cached datasets to the jobs and allow them
to deploy the jobs. 
Hoard integrates transparently with Kubernetes. Datasets are represented
as Kubernetes \emph{custom resources} and they are cached
and made available to deep learning containers as \emph{persistent
volume claims}.

Users inform Hoard about available (remote) datasets by using the standard
Kubernetes administration interface (i.e., \texttt{kubectl}) to create
the \emph{dataset} custom resource which collects dataset meta-data,
a unique dataset name, the URL identifying the remote dataset location, and related access
credentials (we currently support remote stores exposed via NFS or via
S3-compatible APIs~\cite{s3api} for an object store).

Deep learning jobs are also submitted to the system as custom
resources. A \emph{DL job} resource
consists of the training job details such as the number of nodes and GPUs to use, 
the container image to use and additionally specifies the dataset to use by 
specifying its unique name and the path within the container where
the dataset files should be mounted.

Combining the information about compute resource availability (i.e.,
CPU, memory and GPUs) and cache-dedicated storage availability,
and if the dataset is not in the cache, Hoard selects a
set of ``cache-nodes'' for the dataset and another set to run the
training containers. These two sets are co-selected to maximize locality
of containers and cache-nodes, also taking into account the data-center
topology (for example, rack-locality is prioritized if node-locality
cannot be satisfied).  A detailed discussion of the topology optimized
scheduling is outside the scope of this paper. 
Under the hood, the transparent cache will be
automatically mounted inside the training containers at the path specified
by the user. Remote data is fetched either at first access or (optionally)
pre-fetched asynchronously as soon as the dataset resource is created.

Cache eviction is managed at dataset granularity.
When the cache is full, we currently support two eviction options: 
(i) we do not cache new datasets until the user manually specified
one or more existing dataset to evict and (ii) we evict full datasets
on an LRU basis. As the analysis of optimal eviction policies would
require a thorough discussion of its own, we leave it out of the scope
of this paper.

\subsection{Hoard system design and implementation}
\label{sec:impl}

\begin{figure}[t]
	\center
	\includegraphics[width=.8\columnwidth]{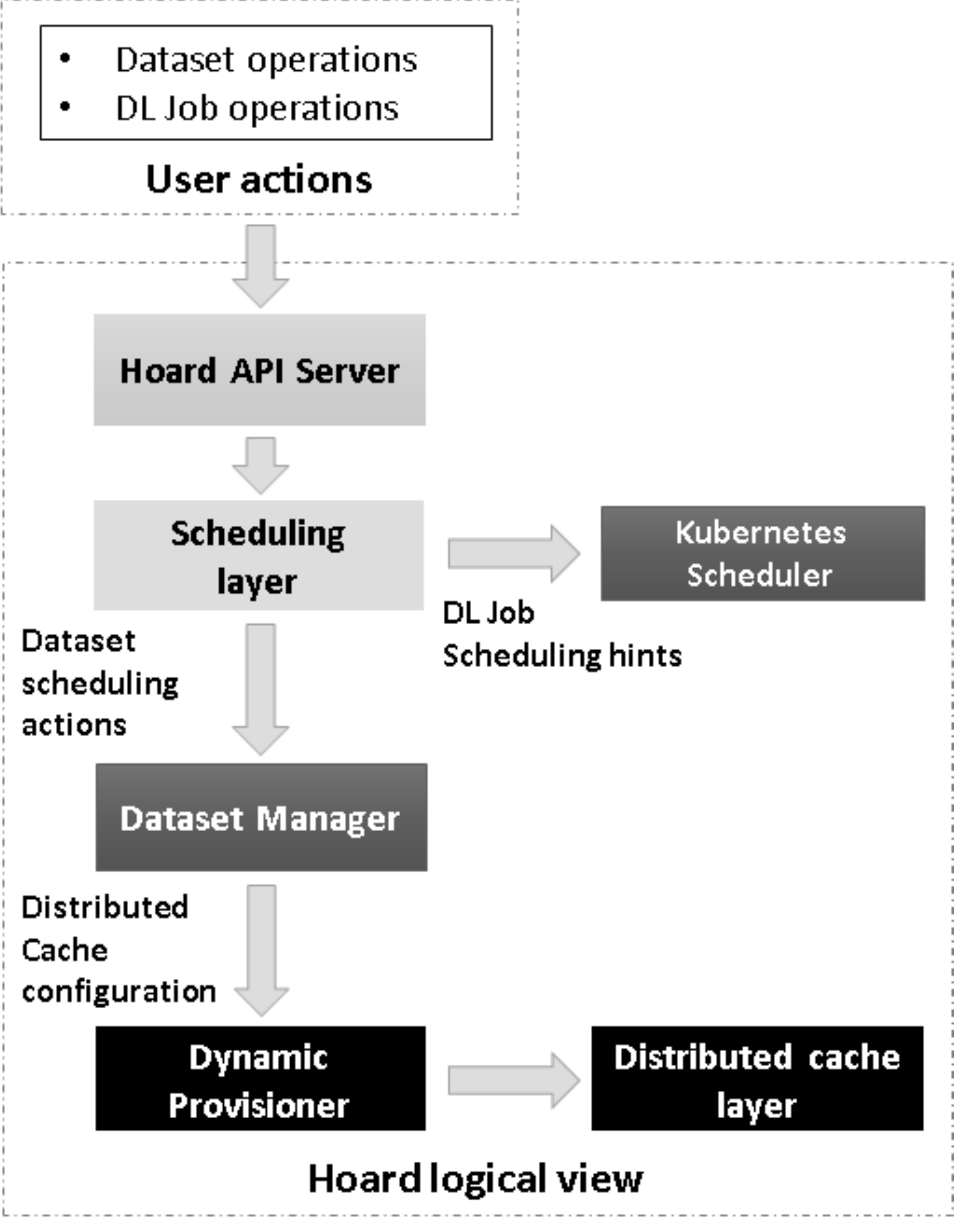}
	\caption{Hoard architecture.}
	\label{fig:impl-arch}
\end{figure}
As shown in the architecture of Hoard \figurename{~\ref{fig:impl-arch}},
it is implemented as a collection of micro-services. 

\emph{The distributed cache layer} at the bottom realizes the
distributed cache on data-center nodes. It implements the transparent
caching logic, handles cache hits and misses and exposes a view of the
full dataset as a POSIX file system in coordination with the distributed cache layer. It accepts command on \emph{what}
and \emph{where} to cache or \emph{eviction} or \emph{pre-fetch},
 but it does not make these choices on its own. The \textit{dynamic provisioner} micro-services creates data volumes
for datasets not present in the cache. 
The \emph{Scheduling layer} at the top is responsible for listening to the creation of
\emph{DL job} and \emph{dataset} custom resources; based on those and
on current resources availability, it makes data and jobs scheduling
decisions.  In the middle of these two layers, sits the \emph{dataset
manager layer}. It acts on the decisions taken
by the scheduling layer, and based on those decisions, it configures
and issues commands to the \emph{dynamic provisioner} and the \emph{distributed cache layer}.

The scheduling layer combines a scheduler service with two Kubernetes
\emph{custom resource controllers} (i.e., one for the \emph{DL job}s
and one for the \emph{dataset}s).  After scheduling decisions are taken
and the underlying cache is configured, the scheduler service encodes
scheduling decisions by using Kubernetes labels and delegates the actual
scheduling of pods to the default Kubernetes scheduler.

The dataset management layer features a dataset-control API service that
accepts commands from the scheduling layer and translates them into
configuration commands for the distributed cache layer. As mentioned above, it also includes
a Kubernetes dynamic volume provisioner that exposes the cached datasets
as \emph{persistent volume claims} once the underlying distributed cache
is set up.

Finally, the distributed cache layer exploits a distributed file system
deployed across the nodes of the data-enter to exploit the storage devices
available on each node as local cache.

\subsection{Selection of file system for the distributed cache}
A distributed file system is at the core of the \emph{Hoard},
so we performed a comparison of some of the most promising and
widely used file systems available today (either open source or proprietary) to satisfy the
requirements stated above and achieve the best performance possible. We
have compared the raw performance in a DL training application of
GlusterFS~\cite{boyer2012glusterfs}, Alluxio~\cite{Li:2014:TRM:2670979.2670985}
and IBM Spectrum Scale~\cite{schmuck2002gpfs}. GlusterFS and Alluxio are open
source, while Spectrum Scale is a proprietary solution.

The benchmark for the comparison is a single epoch training of Resnet50 using 4
GPUs Nvidia P100 and a BS of 128 images per GPU. The results are shown in
\tablename{~\ref{tab:fs-comparison}}. As the table shows, the three file systems enable
a similar training performance. Further analysis of GlusterFS showed that 
it does not support a cache mode out of the box, i.e., it can be used as a cache to 
store the data from another central storage system. We could modify the 
GlusterFS code so it can be a cache but we discarded GlusterFS because of
this limitation.

\begin{table}[t]
	\caption{Comparison of distributed file system solutions for DL training}
	\label{tab:fs-comparison}
	\vskip 0.15in
	\begin{center}
	\begin{small}
	\begin{tabular}{|l|c|}
		\hline
		\textbf{File system} & \textbf{Training duration(min)} \\ \hline
		GlusterFS & 28.9 \\ \hline
		Alluxio & 28.6 \\ \hline
		Spectrum Scale & 27.5 \\ \hline
	\end{tabular}
	\end{small}
	\end{center}
	\vskip -0.1in
\end{table}

Both Alluxio and Spectrum Scale are designed to be configured to cache a remote store, with
Spectrum Scale performing slightly better. A key requirement for the cache (Requirement 1 above)
is that it should allow us to specify a \textit{subset} of the nodes of the cluster to cache a 
particular dataset.
This will constrain the data to a set of nodes and allow for better co-location
of the data and jobs. Unfortunately Alluxio does not allow us to define a set of nodes to 
cache the data, but rather it will use all nodes to cache every dataset. Fortunately, Spectrum Scale
allows for such selection so we 
chose it to implement the cache layer. This
feature is of utmost importance to enable coordinated scheduling of datasets and
DL jobs. Spectrum Scale is deployed on the nodes of the data-center as a kubernetised
service where it exposes a fully POSIX compliant distributed file
system (satisfying our requirement 4).  Active File Management (AFM)~\cite{afm-website} is the extension that
(among other things) allows to use the distributed storage provided by Spectrum
Scale as a transparent cache to remote stores.  Spectrum
Scale~\cite{schmuck2002gpfs} exposes a fully POSIX compliant distributed file
system on shared-nothing clusters.

\section{Evaluation}
\label{sec:evaluation}

\begin{table}[t]
	\caption{Experimental cluster nodes hardware/software configuration}
	\label{tab:hwsw-conf}
	\vskip 0.15in
	\begin{center}
	\begin{small}
	\begin{tabular}{|l|c|}
		\hline
		\multicolumn{2}{|c|}{\textbf{Hardware configuration}} \\ \hline
		\multirow{ 2}{*}{CPU} & IBM Power S822LC dual socket \\
			& (8 cores, 10 threads, NVlink) \\ \hline
		System Memory & 512GB DDR4 \\ \hline
		\multirow{ 2}{*}{Local Storage} & Samsung NVMe SSD 960 Pro \\
					  & (4 x 512GB) \\ \hline
		GPU & 4 x NVidia Tesla P100 \\ \hline
		\multirow{ 2}{*}{Network} & Mellanox ConnectX-5 \\
			    & 100G Ethernet \\ \hline \hline
		\multicolumn{2}{|c|}{\textbf{Software configuration}} \\ \hline
		\multirow{ 2}{*}{OS} & Ubuntu 16.04 \\
		   & (Linux 4.4.0-128-generic) \\ \hline
		\multirow{ 2}{*}{Kubernetes} & v1.9.1+icp-ee \\
		           & (IBM Cloud Private 2.1.0.3) \\ \hline
		IBM Spectrum Scale & v5.0.0 \\ \hline
		Benchmark & AlexNet, BS=1536, 4xGPU \\ \hline
	\end{tabular}
	\end{small}
	\end{center}
	\vskip -0.1in
\end{table}

\begin{figure*}[t]
	\center
	\includegraphics[width=\textwidth]{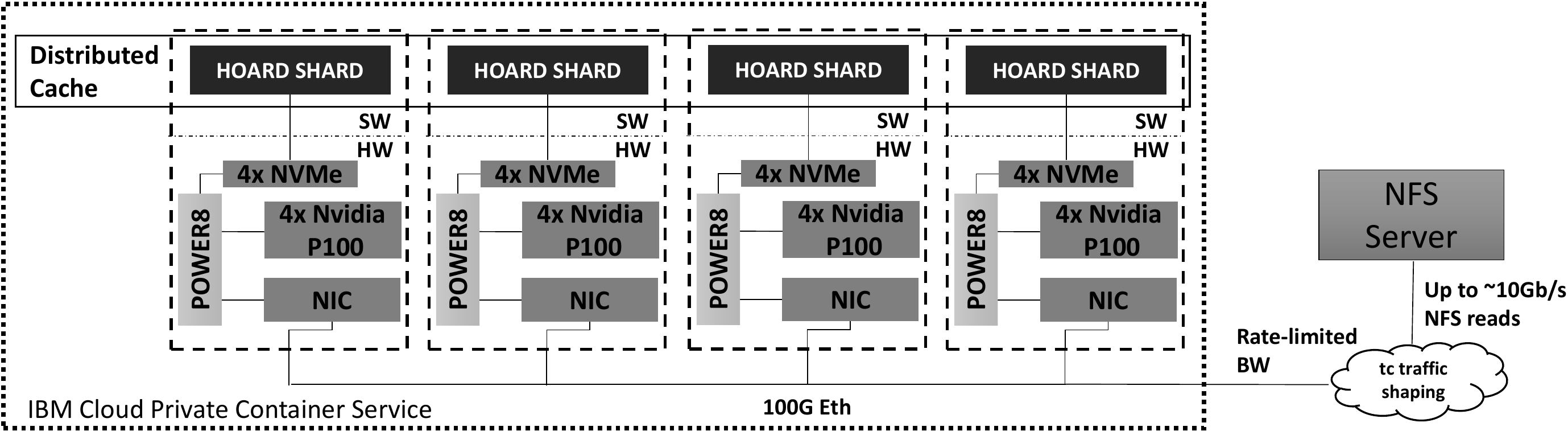}
	\vskip -0.2in
	\caption{Test cluster setup}
	\label{fig:exp-setup}
\end{figure*}

In this section, we evaluate Hoard and demonstrate how it solves the 
problems described earlier.

All the experiments were conducted on a 4 node
POWER8\textsuperscript{\tiny TM} cluster (\figurename{~\ref{fig:exp-setup}})
using NVLink \cite{nvlink-website} connected NVidia P100 GPUs.  All the nodes
feature PCIe-attached NVMe disks and reside on the same 100GbE network (see
\tablename{~\ref{tab:hwsw-conf}} for full node configuration). The system is managed
under an IBM Cloud Private environment~\cite{icp}. We kept our
datasets on a remote NFS server residing on a different network and delivering a
maximum aggregated data bandwidth of ${\sim}1.05GB/s$ when measured from
applications. Although the scale is evidently smaller compared to expected
data center-wide deployments of our system, we argue that this set up is sufficient
to validate our assumptions and to provide insights on our solution.

The experiments use 4 jobs (1 job per node) with 4
GPUs each, training AlexNet with a batch size of 1536 images per GPU. The
training dataset is ImageNet~\cite{Deng09imagenet} and it is ${\sim}144GB$ on disk. The training
script is part of TensorFlow CNN benchmarks suite~\footnote{TensorFlow CNN
benchmarks repository: \url{https://github.com/tensorflow/benchmarks/tree/master/scripts/tf\_cnn\_benchmarks}}. IBM Spectrum
Scale with AFM is installed on each node of the cluster and uses 2 NVMe devices
per node as the back-end for the cache.

Even though our experiments run 4 distinct DL training jobs, the results can be
easily projected to a distributed training. From the perspective of storage the
access throughput needed is driven by the number of GPUs of the job rather than by the
number of nodes used or the number of training jobs in those GPUs.  We run 4 jobs using 4
GPUs, for a total of 16 GPUs being serviced by the same storage; this
could be mapped to a single training job distributed over 4 nodes with 4 GPUs
each. 
A second note is related to the choice of the
deep learning model, AlexNet. The decision was taken by considering the
trade-off between ease of access to the model, reproducibility for community, 
and input requirements (i.e.,number of input elements processed per second). 
That said, it can easily be shown in a distributed multi-accelerator environment, 
the source storage could be stressed in a similar fashion, or more, using 
a large number of worker nodes training ResNet50, e.g. ResNet50 running on 16 
Tesla Volta100 requires 15.5k images per second~\cite{nvidia-hgx-2}.
The TensorFlow CNN benchmark
suite is freely available and designed to measure the performance of a
TensorFlow model. AlexNet, among the other networks available in the suite, is
one requiring higher input data throughput per GPU and thus demanding
in terms of storage. Our approach is not tied to TensorFlow, deep
learning applied to computer vision or AlexNet. Any other network model,
developed for any machine learning framework, with high storage throughput
requirements can immediately benefit from using Hoard, e.g. multi-modal input
sensors, high-fidelity sensors.

In light of the requirements introduced in Section~\ref{sec:requirements}, we
use the experimental section to validate that existing OS based in-memory
caching techniques cannot cope with the storage requirements of a deep learning
training (Requirement 1 \& 2). On the contrary, we expect our approach to be almost
agnostic on the availability of memory on each node of the data-center.  In
addition we also aim at confirming that even though our distributed caching
system is sharing the network with the rest of applications, the amount of
network traffic introduced for coordination of the cache nodes is not
interfering with applications data traffic. Finally, we  give some insights on
how scheduling of DL jobs in relation to datasets placement is affecting the
usage of network at the data-center level (Requirement 3).

\subsection{Baseline cache performance compared to remote and local storage}
\label{sec:cache-effect}

\begin{figure}[t]
	\center
	\includegraphics[width=\columnwidth]{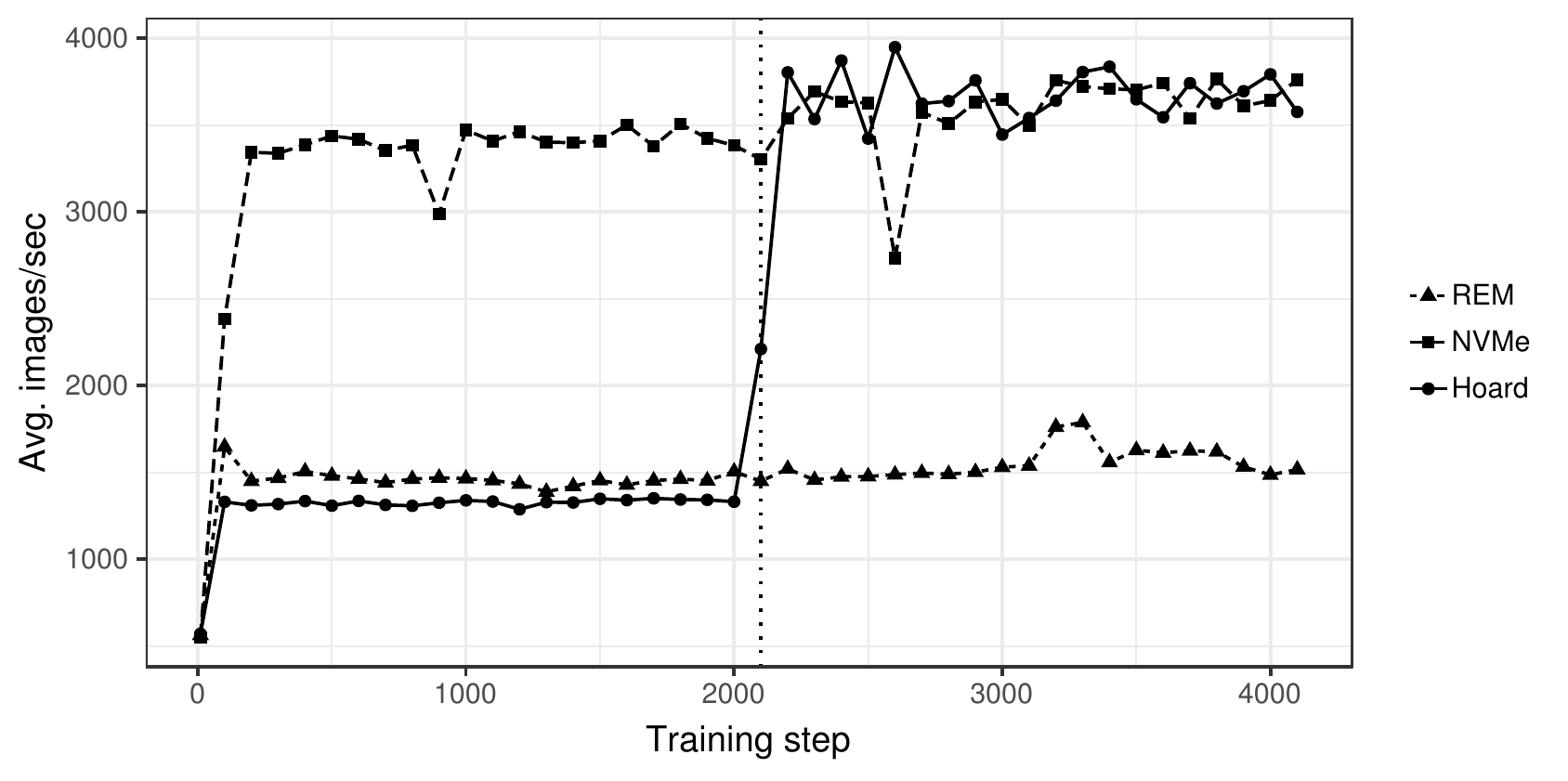}
	\vskip -0.1in
	\caption{Training performance during a two epochs training. The vertical
	line represents the boundary between first and second epoch}
	\label{fig:exp-cache-effect}
\end{figure}

First we needed to ensure that our cache can perform as well as the remote store for epoch 1
and as well as the local disk for subsequent epochs.
In this experiment our AlexNet benchmark 
is executed for a total of 2 epochs with remote storage alone, local disk alone and with Hoard.

\figurename{~\ref{fig:exp-cache-effect}} compares the two solutions currently
applied to deep learning in data-centers and our Hoard caching solution: reading data directly from remote
storage (REM) and copying the dataset to the local storage (NVMe, SSD, etc) of each node  before
starting the DL Jobs (NVMe). As expected, in the first $2000$ training steps, the NVMe case is significantly faster
as data is accessed from local storage.
Hoard performs as good as the remote store for the first epoch and as good as the local disk for epoch 2. 
This establishes that the Hoard does not impose any overhead on the application performance.
But now the entire cluster storage is available for any single job needing up to 4TB of input data (each node has 1TB NVMe cache) 
where it was limited to 1TB without Hoard.
In addition, Hoard manages the life cycle of the data in the cache so if the job
is executed multiple times with think-time or with different hyper parameters, the cached data will be used, instead
of fetching the data from the remote store.

\begin{table}[t]
	\caption{Long training speedup projections with remote storage as baseline}
	\label{tab:exp-speedup}
	\vskip 0.15in
	\begin{center}
	\begin{small}
	\begin{tabular}{|l|c|c|c|c|}
		\cline{2-5}
		\multicolumn{1}{c|}{} & \textbf{2 epochs} & \textbf{30 epochs} &
		\textbf{60 epochs} & \textbf{90 epochs} \\ \hline
		REM  & $1~\times$ & $1~\times$ & $1~\times$ & $1~\times$ \\ \hline
		Hoard & $0.93~\times$ & $1.98~\times$ & $2.07~\times$  &
		$\bm{2.1~\times}$ \\ \hline
		NVMe & $2.28~\times$ & $2.3~\times$ & $2.32~\times$  &
		$2.32~\times$ \\ \hline
	\end{tabular}
	\end{small}
	\end{center}
	\vskip -0.1in
\end{table}

If we project the performance of the figure over a long training
(\tablename{~\ref{tab:exp-speedup}) we can conclude that \emph{Hoard achieves a 
2.1x improvement in execution over the shared storage}.
The slower first epoch would not significantly
impact the overall training time especially considering hundreds of epochs are
often required during neural network training. Because Hoard places the data close to the compute,
the jobs finish 2x faster that means at least 2x more jobs can completed with Hoard in the time
1x jobs complete with the shared storage. 
As the cluster size grows and it can run more jobs, these jobs will place bigger load on the shared storage
so Hoard's advantage over shared storage grows as well. 
\textit{Thus Hoard improves the utilization of the cluster by 
at least 2x.}

\subsection{Impact of system memory availability on the efficacy of Hoard cache}
\label{sec:exp-sysmem}

\begin{figure*}[t]
	\center
	\includegraphics[width=0.8\textwidth]{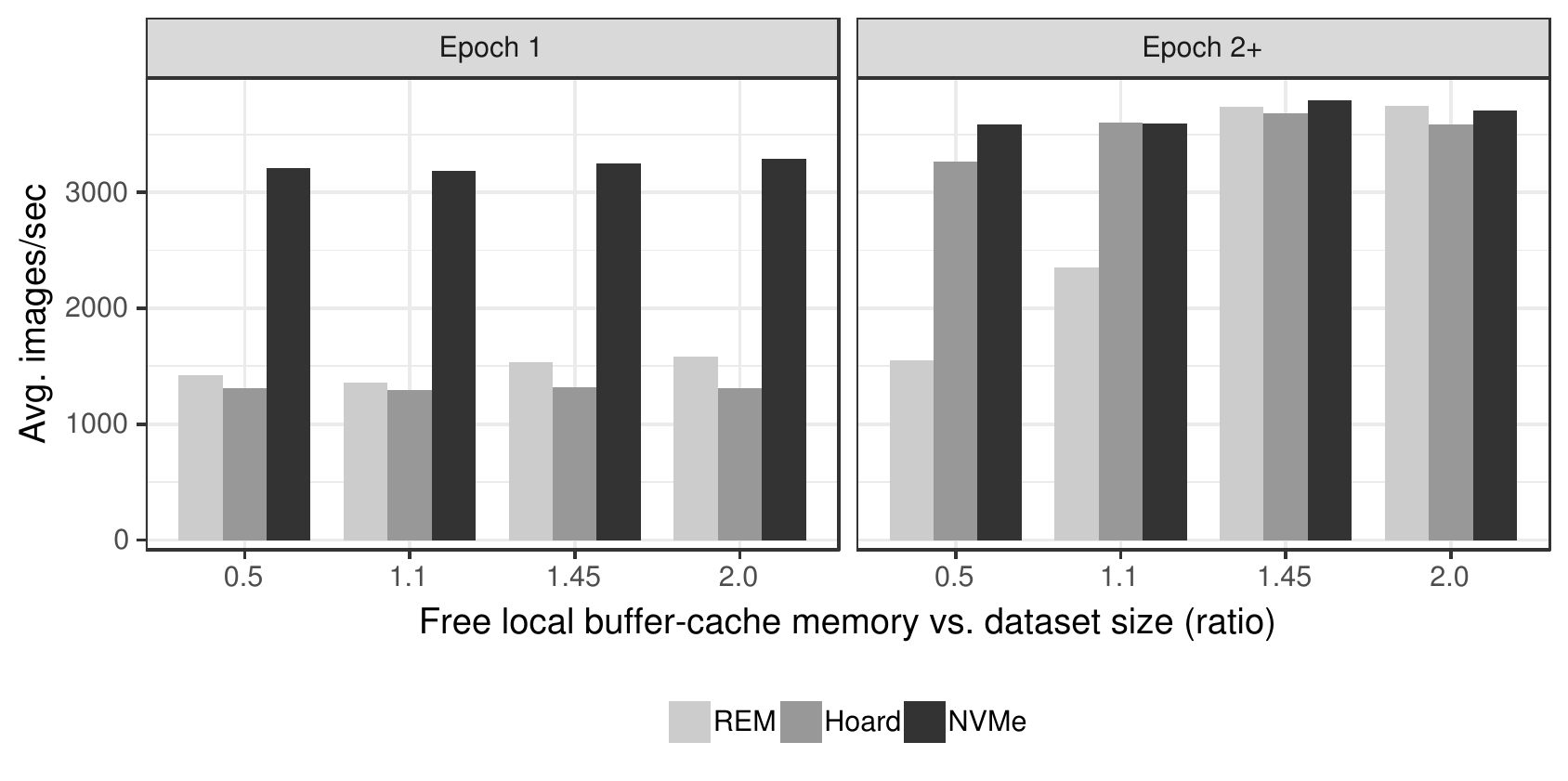}
	\vskip -0.3in
	\caption{Training performance with multiple ratios of available system
	memory vs. size of the dataset, for first and subsequent epochs}
	\label{fig:exp-sysmem}
\end{figure*}

The previous section does not take into account possible memory-based caching done by
the operating system that could improve the performance while accessing the
datasets directly from the remote storage.

The most common memory caching technique is the Linux
buffer-cache~\cite{buffer-cache} that
opportunistically caches frequently accessed files blocks (1KB is the default
block size) into the free memory available. Having the whole dataset cached in
memory would make every access being serviced from DRAM rather than network or
local storage, greatly improving the storage throughput perceived by
applications.  In this experiment we evaluate the impact of OS memory-based
caching by varying the ratio between free memory available on each of the nodes
and size of the dataset used (or MDR for brevity).
Hoard does not benefit from OS buffer cache because it uses a portion of memory
dedicated to Spectrum Scale (\emph{pagepool}). In the case of buffer cache the
MDR is changed using the Linux \emph{stress} tool \footnote{Linux stress command man
page: \url{https://linux.die.net/man/1/stress}} that can be programmed to spawn a process allocating an amount of memory
defined by the user. The stress process will play the role of third party
applications using an arbitrary amount of memory, or other training jobs
co-located on the same nodes and thus sharing the system memory available.
In case of Hoard we set the size of the Spectrum Scale pagepool to reach the desired MDR
value.

\figurename{~\ref{fig:exp-sysmem}} shows the average training performance, in
frames per second (fps), with different MDR values.
When the MDR is $>~1.1$, the whole dataset is cached in memory after the first
epoch and the three solutions show the same performance. In a real deployment
this is an unlikely scenario as it would require enough memory to host a full
dataset and should also consider memory used by other applications. However,
it is still useful to show the effectiveness of in-memory caching. In all the
following experiments we fix the MDR to $0.5$ as it represents a more plausible,
yet generous, scenario.

Reducing the MDR has a detrimental effect on the REM case, as more data starts
being fetched from remote storage during all the epochs of the training. The OS 
likely starts trashing the buffer cache bringing in file blocks
recently read from NFS evicting others that will be needed in the near future
due to the repetitive nature of DL training. It is interesting to see how Hoard
is almost completely agnostic to the amount of memory assigned for in-memory
caching (\emph{pagepool}) by still being able to deliver the same performance of
local NVMe storage access (the one in the figure for NVMe during first epoch)
even when the MDR is set to $0.5$.
For the NVMe case even the minimum amount of buffer cache available is
beneficial as it adds on top of the already high performance of NVMe devices.

\emph{This second experiment reveals that not only our approach utilizes the storage
more effectively, it also benefits from free DRAM that is used for caching}.
This translates into a higher potential degree of multi-tenancy as the
data-center management might decide to dedicate a small amount of
memory to the Spectrum Scale pagepool and co-locate on the same node DL Jobs,
heavily using GPUs, along with CPU and memory intensive jobs.

\subsection{Impact of Hoard caching on remote storage bandwidth}
\label{sec:exp-rembw}
\begin{figure*}[t]
	\center
	\includegraphics[width=0.8\textwidth]{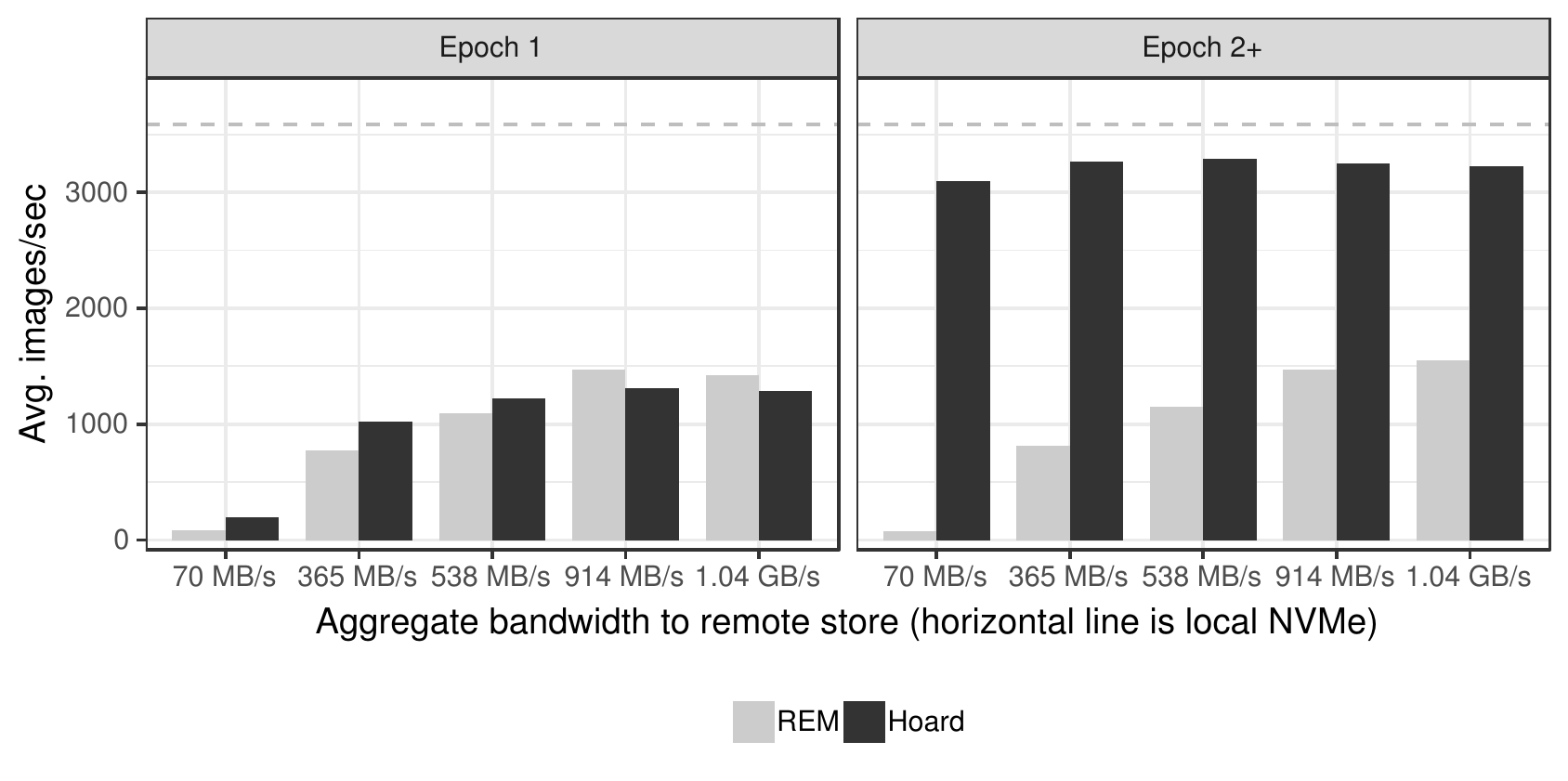}
	\vskip -0.3in
	\caption{Training performance with multiple values of remote storage
	bandwidth, for first and subsequent epochs}
	\label{fig:exp-rembw}
\end{figure*}

Different users might use different remote storage solutions that mainly differ
in their read/write bandwidth. The NFS server used in our experiments allows a
maximum read bandwidth measured from application of $1.05GB/s$ (used for all
previous experiments) that could be already representative of off-the-shelf
cloud storage.
However for the sake of better understanding the
characteristics of our framework we conducted an experiment where the NFS read
bandwidth is further scaled down.
To decrease the remote storage bandwidth perceived by applications we act on the NFS
server (\figurename{~\ref{fig:exp-setup}) using the Linux traffic control
tool\footnote{Linux tc command man page: \url{https://linux.die.net/man/8/tc}}
that allows setting limits on the bandwidth of a network interface.

The results in \figurename{~\ref{fig:exp-rembw}} validate an already expected
behavior. While accessing training data directly from remote storage is heavily
dependent on the bandwidth, Hoard is linked to it only during the first
epoch. Subsequent epochs get close to local storage access throughput and
the performance of the training job goes again to its maximum.
\emph{Our approach performs equally better regardless of the remote storage solution
selected by the user}.

\subsection{Impact of Hoard cache on data-center network usage}
\label{sec:exp-nwintf}

\begin{table}[t]
	\caption{Network usage during training}
	\label{tab:exp-bw}
	\vskip 0.15in
	\begin{center}
	\begin{small}
	\begin{tabular}{|l|c|c|c|}
		\cline{2-4}
		\multicolumn{1}{c|}{} & \textbf{Total data} &
		\textbf{Transmission} & \textbf{Training}\\
		\multicolumn{1}{c|}{} & \textbf{transmitted} & \textbf{rate} &
		\textbf{duration}\\
		\multicolumn{1}{c|}{} & \textbf{(TB)}  & \textbf{(Gb/s sent)} &
		\textbf{(hours)}\\ \hline
		REM  & $8.1$ & $1.23$ & $14.90$ \\ \hline
		Hoard & $8.1$ & $2.7$ & $6.97$ \\ \hline
	\end{tabular}
	\end{small}
	\end{center}
	\vskip -0.1in
\end{table}

Although an individual job's performance is incredibly important, we must consider
overall throughput of the entire system in a multi-tenant environment. Consider the
impacts on network communication when using the distributed cache where this
overhead could potentially degrade the overall performance because of the control messages
associated with the distributed cache.

To understand where we stand with respect to network usage we monitored, during 60 epochs of
training, the amount of data exchange rate sustained by each node of the Hoard
cluster while communicating with its peers, together with the total amount of
data exchanged. We performed the same monitoring also in the case when datasets
are directly accessed from remote storage. We also measured the duration of the
training itself. The results collected are reported in
\tablename{~\ref{tab:exp-bw}} and show the average network traffic generated for 1
training job using 4 GPUs.

As a first validation, the total amount of data transmitted over the network is
matching in both the cases considered (${\sim}8TB$, roughly $144GB \times 60$ epochs).
The second interesting value to observe is the transmission rate. In the Hoard
case it is the aggregate bandwidth sent by each node to all its peers, averaged
over the whole training. While for the REM case the value is the average data 
sent rate by the NFS server for each of the four jobs running.
The data transmission rate is ${\sim}2.1\times$ higher in the
Hoard case. This results is somewhat expected as the training took ${\sim}2.1\times$
less than the REM case. We can easily conclude that the higher network bandwidth
usage is not due to any extra communication introduced by GPFS/AFM, but rather
to a faster training where the GPUs were able to process a higher number of
frames per second.
There is minimal traffic generated to coordinate the cache nodes but
it is, negligible and impossible to appreciate because hidden by the huge traffic
generated to exchange the actual dataset data.

\subsection{Do we need to co-schedule data and compute?}

What if
the DL Jobs are scheduled on nodes where data is not locally cached? This is a
possibility and our distributed cache layer supports it.
We have performed a simple experiment where datasets are cached on only two of
the nodes in our cluster (recall Hoard support this), and compared two scenarios, if the jobs are scheduled 
on those nodes versus being scheduled on nodes where the data is not locally cached. 
Due to the scale of our
test cluster, coupled with the storage bandwidth requested by AlexNet, we
could not stress our cache enough and appreciate any significant
difference in performance.
Nonetheless we want to give the reader some insights on what could be the effect
of scheduling of DL jobs and datasets at the data-center level.

\begin{table}[h]
	\caption{Percentage of rack up-link bandwidth (40G network) used by DL
	jobs that are scheduled on a rack where data is not cached}
	\label{tab:exp-uplink}
	\vskip 0.15in
	\begin{center}
	\begin{small}
	\begin{tabular}{|l|c|c|c|c|}
		\cline{2-5}
		\multicolumn{1}{c|}{} & \multicolumn{4}{|c|}{\textbf{Percentage of jobs
		misplaced}} \\
		\cline{2-5}
		\multicolumn{1}{c|}{} & $20$ & $40$ & $60$ & $80$ \\ \hline
		up-link BW & $5\%$ & $9\%$ & $13\%$ & $17\%$ \\ \hline
	\end{tabular}
	\end{small}
	\end{center}
	\vskip -0.1in
\end{table}

We have performed a rack-level analysis of the up-link bandwidth utilization
in relation to the number of DL jobs that fetch data from a dataset cached on a
different rack. The data-center model considered in this analysis is composed of
a number of racks, each of which is equipped with a top of rack switch (TOR) with
32 ports and an over-subscription ratio of 3:1. The TOR switch ports work at
40G with an aggregated up-link bandwidth of 320Gbps

In \tablename{~\ref{tab:exp-uplink}} we show the projection of the percentage
of the rack up-link bandwidth used by a total of 24 DL jobs with a certain
percentage of such jobs scheduled on a rack different from the ones where
their datasets are cached.
The higher the percentage of misplaced jobs in
a rack, the higher is the portion of up-link bandwidth used to access the
datasets and the lower is the bandwidth remaining for other applications
potentially running on the same racks.  The numbers in the table 
do not seem worrisome ($5\%$ of total up-link for $20\%$ of jobs misplaced). So, the co-scheduling
may not be necessary for small scale cluster with substantial backbone bandwidth.
However, newer GPUs deliver already up-to $3\times$ higher
performance in deep learning applications than an NVidia P100 \cite{nvidia-v100} and
future GPUs and accelerators are expected to further improve the performance of
a DL training. A wise scheduling policy would first try to schedule jobs and
datasets on the same node, and if not possible prefer scheduling on the same
rack to avoid network interference with other applications. We speculate that
rack-aware scheduling might be sufficient to achieve a balance between locality,
performance, and flexibility in placement.

\section{Prior Art}
\label{sec:prior-art}

Caching network based file systems is not a new topic in computer science
\cite{nelson1987caching}. There have been multiple works since then using both
Solid State Disks (SSDs) or spinning disks as a local caches, and considering
both single node and distributed deployments
\cite{bent2002flexibility,makatos2010using,li2014nitro,
Saxena:2012:FLC:2168836.2168863,byan2012mercury,ernst2001dcache,eshel2010panache}.

The unique access patterns and usage models of DL applications
offer new opportunities to design distributed caches to improve resource utilization and 
performance of these applications.
Caching remote storage to target I/O bound Deep Learning training applications at
a large scale is instead a relatively newer topic, but there are already a few
related projects that are worth mentioning. 
The first is the Kubernetes Volumes Controller (KVC)~\cite{intel-kvc}.
KVC shares some of the goals of Hoard, i.e., caching
of data in local nodes and co-location of DL training jobs with datasets, but it
does not rely on a distributed cache layer. In case of distributed jobs the
dataset is replicated on each of the nodes involved in the computation, clearly
wasting the local storage available on each node. 
The second interesting approach is
\cite{nvidia-alluxio}, where the Alluxio file system is used as a cache for
datasets stored on a remote S3-based bucket.  Although Alluxio works as a
distributed cache for a remote store, it does not allow for placing datasets on specific nodes.
In addition, the document cited only considers the
perspective of the cache, giving no insights on the user experience.
Hoard is instead a turnkey solution that can be directly used by a data
scientist. Users can in fact interact with Hoard as with any other cloud service
they are familiar with, where no choices are to be made in terms of location of
data, scheduling of applications on compute nodes, etc. The last interesting
approach is \cite{riseml}, where datasets stored on a remote NFS server are
cached on the local storage of a node using the Linux
\emph{cachefsd}~\footnote{cachefsd man page:
\url{https://linux.die.net/man/8/cachefilesd}}. Similar to the first
approach discussed, cachefsd uses storage at the single node level and again in
case of a distributed training job the cached data would be replicated on each
node. In addition, data cached with this mechanism is to be considered volatile,
as the cache is associated with the single NFS mount. Remounting the same
location would re-create the cache. This is clearly a limitation as at the base
of our approach we have the decoupling of datasets and DL Jobs life-cycles, to
enable re-usage of already cached datasets and speedup the deployment of
``returning'' jobs.

To the best of our knowledge, Hoard is the first attempt of a cloud oriented
storage acceleration system targeting I/O bound deep learning training
applications.

\section{Conclusions}
In this paper, we show that the distributed caching as a middleware layer in deep learning systems
can be used to feed the accelerators as fast as they consume the data.
Hoard distributed caching is built using an existing distributed file system 
for caching the data and uses a collection of micro-services that provide
the functions to create caches on subsets of DL system nodes, coordinate with the
job scheduler to co-locate cache and DL training jobs on sets of nodes and manage the
cached data life cycles independently of the job life cycles. This way the cached data
could be used between different invocations of the same job and between jobs that
use the same datasets. We evaluated Hoard on a moderately sized cluster 
with 4 nodes and 16 GPUs and showed that it improves the system utilization by 2x. 
Our implementation allows for scaling Hoard to much larger scale systems and we expect
even higher utilization improvements because of the increased load that would 
be placed on the shared central storage system which will be mitigated by Hoard.  
We believe that deep learning workload-aware caches like Hoard play an increasingly important
role in enabling deep learning workloads on the cloud
because they can bridge the growing I/O gap between the central storage systems and 
the faster accelerators combined with growing datasets.
 
\newpage
\bibliography{bibliography}
\bibliographystyle{sysml2019}

\end{document}